# A DEEP LEARNING NETWORK FOR THE CLASSIFICATION OF INTRACARDIAC ELECTROGRAMS IN ATRIAL TACHYCARDIA

Zerui CHEN[1], Sonia Xhyn TEO[1], Andrie Ochtman[1], Shier Nee SAW[2], Nicholas CHENG[1], Eric Tien Siang LIM[3], Murphy LYU[4], Hwee Kuan LEE[1,5,6,7,8,9]

1. Bioinformatics Institute, Agency for Science, Technology and Research (A*STAR), Singapore, Singapore.
2. Department of Artificial Intelligence, Faculty of Computer Science and Information Technology, Universiti Malaya, 50603 Kuala Lumpur, Malaysia.
3. Department of Cardiology, National Heart Centre Singapore, Singapore, Singapore.
4. Abbott Medical, Singapore, Singapore
5. School of Computing, National University of Singapore, Singapore, Singapore.
6. Image and Pervasive Access Lab (IPAL), Singapore, Singapore

7. Singapore Eye Research Institute (SERI), Singapore, Singapore.

8. Rehabilitation Research Institute of Singapore, Singapore, Singapore.

9. Singapore Institute for Clinical Sciences, Singapore, Singapore.

Address for correspondence:

Nicholas Cheng, Ph.D.,

Bioinformatics Institute, Agency for Science, Technology and Research (A*STAR), Singapore

30 Biopolis Street, #07-01 Matrix, Singapore 138671

Tel: +65 6478-8387; E-mail: Nicholas_Cheng@bii.a-star.edu.sg

# Abstract

A key technology enabling the success of catheter ablation treatment for atrial tachycardia is activation mapping, which relies on manual local activation time (LAT) annotation of all acquired intracardiac electrogram (EGM) signals. This is a time-consuming and error-prone procedure, due to the difficulty in identifying the signals' activation peaks for fractionated signals. This work presents a Deep Learning approach for the automated classification of EGM signals into three different types - "normal", "abnormal", and "unclassified", which forms part of the LAT annotation pipeline, and contributes towards bypassing the need for manual annotations of the LAT. The Deep Learning network, the CNN-LSTM model, is a hybrid network architecture which combines convolutional neural network (CNN) layers with long short-term memory (LSTM) layers. 1452 EGM signals from a total of 9 patients undergoing clinically-indicated 3D cardiac mapping were used for the training, validation and testing of our models. From our findings, the CNN-LSTM model achieved an accuracy of 81% for the balanced dataset. For comparison, we separately developed a rule-based Decision Trees model which attained an accuracy of 67% for the same balanced dataset. Our work elucidates that analysing the EGM signals using a set of explicitly specified rules as proposed by the Decision Trees model is not suitable as EGM signals are complex. The CNN-LSTM model, on the other hand, has the ability to learn the complex, intrinsic features within the signals and identify useful features to differentiate the EGM signals.

# Introduction

Tachycardia is a class of heart arrhythmia where the heart beats at an abnormally fast rate of more than 100 beats per minute while the patient is at rest [1], [2] and this may lead to serious complications such as heart failure, stroke or sudden cardiac arrest if left untreated [3]. One type of tachycardia is atrial tachycardia, where the heart's electrical impulse abnormally originates from an ectopic site in the atria, instead of the normal origin of electrical activity, the sinoatrial node. The underlying basis of atrial tachycardia is the manifestation of continuously circulating electricity where a region of the heart is repetitively excited [2].

Atrial tachycardia may be effectively and permanently treated via catheter ablation, where a specific target location, known as the isthmus, is identified from cardiac electrical mapping and ablated by heat to break the abnormal reentry conduction circuit [4]. Cardiac electrical mapping involves the insertion of intracardiac electrodes to measure the electrical activity on the internal walls of the atria. Using the electrodes, sequential electrical activations of tissue within the cardiac chamber of interest are obtained, and are reconstructed to give the Local Activation Time (LAT) map. The LAT map is constructed via LAT annotations made on the gathered electrogram (EGM) signals, where each LAT annotation represents the point in time when the myocardium underneath the electrode is activated. From the LAT map, the pattern of propagation of electrical waves within the cardiac chamber can be visualised, allowing the ablation target to be identified [4]. This process is straightforward when EGM signals are simple and discrete, with each instance of activation marked by a distinct peak. However, the EGM signals captured near potential ablation sites are mostly complex and fragmented, often with activation peaks that are not well-defined, resulting in ambiguity and errors in annotating the LAT. The errors in the LAT annotations propagate to the LAT maps which may ultimately lead to inaccuracies in the identification of ablation targets. Therefore, the correct classification of the EGM signals according to their types, prior to performing LAT annotations, is a key step in the accurate identification of ablation targets. More specifically, knowing the type of EGM signals (simple and discrete, or complex and fractionated) as the first step will allow us to further quantify the errors made in the LAT annotations, and consequently, those made in the identification of ablation targets.

The use of machine learning in cardiac electrophysiology and arrhythmia has been a topic of intense research, and a more detailed exposition can be found in a review by Trayanova et al. [5]. Most of the applications of machine learning are based on the analysis of time-series recordings, with the most commonly used signal being the body-surface electrocardiogram (ECG) [5], and there were relatively fewer studies applying machine learning techniques on intracardiac electrogram signatures obtained within heart chambers, the subject of this study. Also, application of machine learning is much focused on atrial fibrillation, another class of tachycardia, and there is a paucity of studies performed on atrial tachycardia. Smith et al. developed a deep neural network for 12-lead ECG interpretation for the diagnosis of atrial fibrillation [6]. Hannun et al. [7] used a deep learning network for the classification of 12 rhythm classes using single-lead ECGs from patients using a single-lead ambulatory ECG monitoring device for diagnosing arrhythmia. McGilivray et al. adopted a Random Forest classifier that used 24 EGMs features to identify re-entrant drivers in a simulated model [8] and showed 95.4% accuracy. Muffoletto et al. demonstrated a proof-of-concept showing that deep learning models were able to predict optimal ablation strategies using the simulated data from their simulation with 79% accuracy [9]. Another study used a fuzzy-decision tree to classify the EGMs in four different classes of Complex Fractionated Atrial Electrograms (CAFE) [10] and showed an accuracy of 81% [11]. Duque et al study showed that K-nearest neighbour classifier achieved an accuracy 92% using 92 hand-crafted features extracted from EGMs [12]. To the best of the authors' knowledge, and at the point of writing, there is no study performed on the application of machine learning on analysing intracardiac EGM signals from atrial tachycardia.

To address the lack of research in the analysis of EGM signals arising from atrial tachycardia, the focus of this paper is the automated classification of intracardiac EGM, which forms a crucial step in the process of locating the isthmus for ablation in atrial tachycardia. The types of intracardiac EGMs observed give an indication of the state of health and disease within the conducting cardiac tissues. In this study, we hypothesise that the classification of intracardiac EGMs can be effectively automated with machine learning. We have developed a Deep Learning-based model, a hybrid convolutional neural network with long short-term memory cells, for this task. The proposed model showed promising results in achieving an overall 81% one-vs-all classification accuracy on our clinical dataset. For comparison, we have separately developed a rule-based model generated by using *a priori* knowledge gathered from visual observation - the Decision Trees model - which achieved an overall 67% one-vs-all classification accuracy on the same clinical dataset. Our results show that the deep learning model outperformed the rule-based model. In all, we have demonstrated the feasibility of using an automated deep learning-based approach for the classification of EGM in tachycardia.

# Methods

## *Study Design*

In this retrospective study, we included a total of 9 patients undergoing clinically-indicated 3D cardiac mapping and ablation at the National Heart Center Singapore (NHCS). An average of 160 EGM signals were gathered per patient, amounting to a total of 1452 EGM signals. An example of an EGM signal is shown in Fig. 1, and can be thought of as a time series of voltage values with interspersed periodic signal peaks indicating cardiac activations. EGM signals are accompanied by their mapping data, which indicates the spatial location within the cardiac chamber where each signal is collected, and together, these allow a 3D mapping of the electrical activity of the heart to be reconstructed.

Ethical approval was obtained from SingHealth Centralised Institutional Review Board (CIRB) for this study. All participants gave their informed consent prior to their enrollment for this study. The collection of intracardiac EGMs was performed in accordance with relevant guidelines and regulations. All collected data in this study had been anonymized and were used solely for research purposes. The results of this study were not used to direct the clinical treatment of the participating patients.

## *Electrophysiological Study and Data Collection*

The 3D cardiac mapping EGM signals were collected by a trained electrophysiologist with catheters (10 polar, Biosense Webster, Diamond Bar, USA) using a standard 3D anatomic mapping procedure [13], [14].

## *EGM Classification*

Three distinct classes of EGM signals were identified in this study (Fig. 1): the "normal" type, which indicates propagation in normal cardiac tissues (Fig. 1a); the "abnormal" type, which characterises propagation patterns near abnormal tissues (Fig. 1b); and a third, "unclassified" type, which is used to label EGMs that present a high degree of uncertainty and do not neatly fall under the "normal" or "abnormal" types (Fig. 1c). "Unclassified" EGM signals could be attributed to the non-ideal contact between the measuring electrode and tissue, or they could also be due to weak electrical activity generation from the tissue. This EGM categorisation was derived based on a previous study by Frontera et al. (2018) [15]. The classification of the EGM signals is of significant clinical importance because abnormal signals comprise fractionated signals that are commonly seen near the electrical reentrant point, slow conduction region and line of block. Thus, good classification of the EGM signals can help in the development of an accurate LAT map, which will result in better identification of the isthmus for ablation.

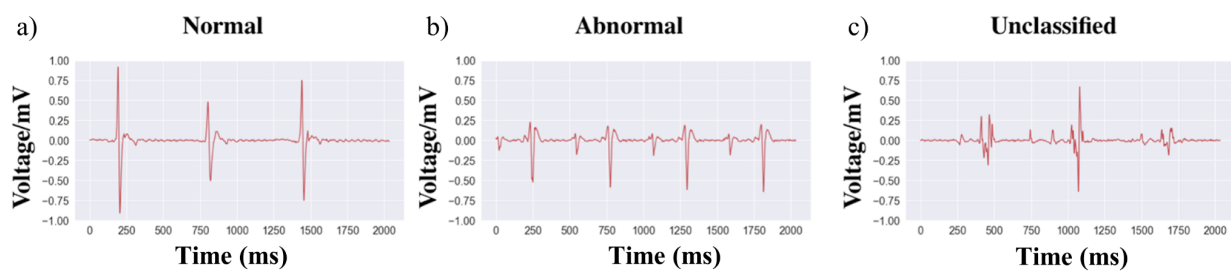

*Figure 1. Representative examples of normal, abnormal and unclassified EGM signals. (a) shows an example of a 'normal'-labelled EGM signal, with a distinct signal peak indicating tissue activation. (b) shows an example of an 'abnormal'-labelled EGM signal, with a 'fragmented' characteristic showing many peaks in close proximity, with amplitude smaller than normal signals. (c) shows an example of EGM deemed to be 'unclassified', which exhibits signals of high amplitude, as well as a certain degree of fragmentation (which could be due to signal noise).*

## Data Split and Annotation Strategy

The entire dataset consisting of nine patients' data was randomly split into training, validation and testing subsets. Seven patients' data were chosen for the training subset, one patient's data for the validation subset and the remaining one patient's for the testing subset. Each patient's data consists of multiple EGM signals, as shown in Table 1.

Two electrophysiologists and three AI researchers were involved in the annotation of the EGMs into the above-mentioned types - 'normal', 'abnormal' and 'unclassified'. A subset of the test dataset (consisting of 1 patient) with a total of 135 EGM signals (comprising 60 normal, 37 abnormal, and 38 unclassified signals) was annotated by the two electrophysiologists, while a strategy of unanimous annotations performed by the AI researchers was adopted for the annotation of the remaining test, training and validation datasets. Prior to the AI researcher's annotation task, to ensure the quality of annotations performed by the AI researchers, the electrophysiologists had demonstrated the process of EGM signal annotation to the AI researchers using a subset of the test dataset. Each AI researcher then individually carried out the annotation on the same subset of the testing dataset and every signal was given a label based on the unanimous agreement of all three researchers, in other words, each signal was only given a label agreed upon by all 3 researchers. The rationale of this was also to minimise interobserver variability. The researchers' unanimous annotations were then compared against the corresponding annotations performed by the two electrophysiologists to ensure labelling quality. The results for this comparison are shown in the Results section. This strategy was used in the annotation of the rest of the testing, training and validation dataset, and only EGM signals with unanimous labels are considered for model training and validation. Non-unanimously-labelled EGM signals were discarded. This effectively reduced the total number of signals from a total of 1452 to 820, after filtering out the signals in which the three researchers produced different labels. The breakdown of the resulting dataset is tabulated in Table 1.

*Table 1.* *Overview of unanimously-labelled EGMs by AI researchers and the number of signals used in training, validation and testing datasets*

|  |  | Data Split | | |
|---|---|---|---|---|
|  |  | **Training** | **Validation** | **Test** |
| **Number of Patients** | | 7 | 1 | 1 |
|  |  |  |  |  |
| **Type of signals** | **Normal EGM** | 199 | 44 | 83 |
| | **Abnormal EGM** | 138 | 25 | 52 |
| | **Unclassified EGM** | 198 | 38 | 43 |
| | **Total EGM** | **535** | **107** | **178** |

## Data Augmentation

Data augmentation was performed to increase the variance of the dataset. First, random cropping was applied to the input EGMs first (Fig. 2a), where random segments of window width 1500ms were cropped out and discarded. This step increased the variation of our dataset by generating more unique training samples to mitigate overfitting. After random cropping, the EGMs were then normalised by linear scaling to have a maximum absolute amplitude of one (Fig. 2b).

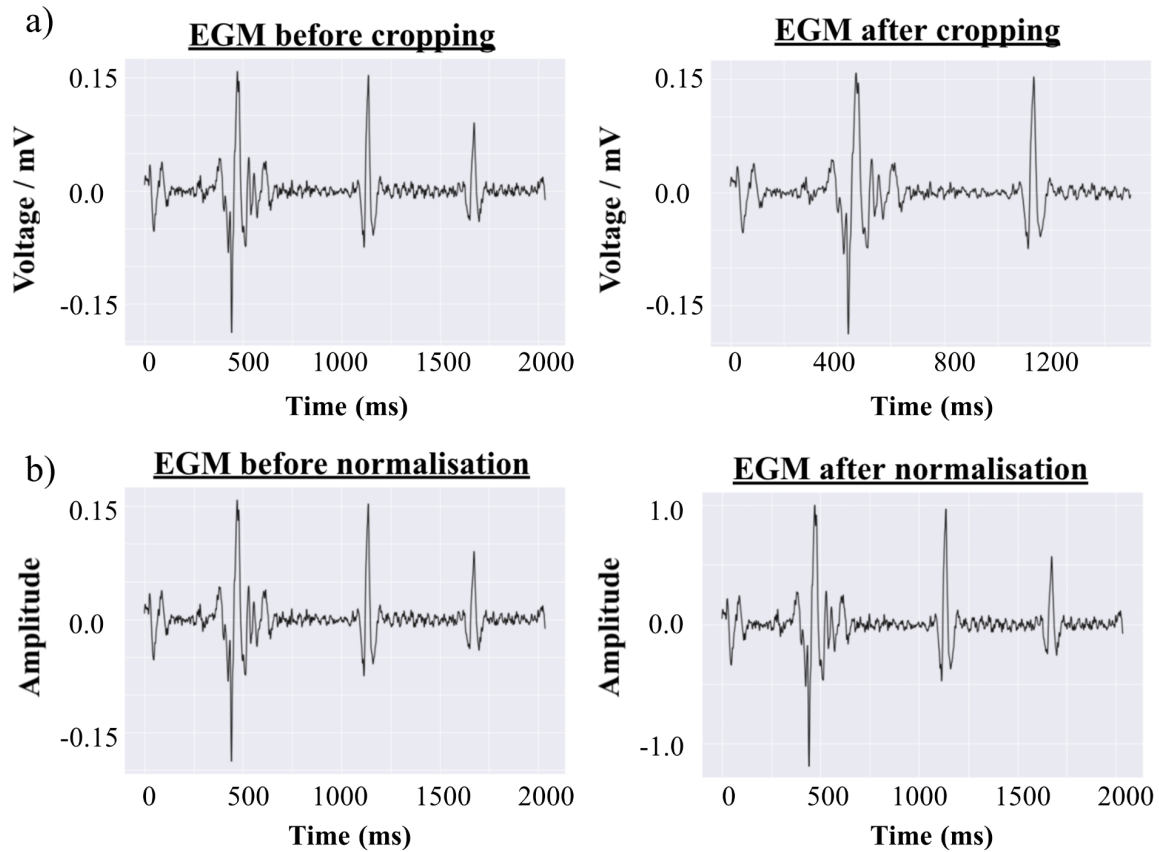

***Fig. 2.*** *Data Augmentation (a) The effect of random cropping with a window length of 1500ms. The boxed region from the original EGM (left) was cropped out to give the augmented EGM (right). (b) The effect of normalisation. The original EGM (left) is normalised to have a maximum absolute magnitude of 1 (right).*

## *Deep Learning Model: Convolutional Neural Network, with Long Short-term Memory (CNN-LSTM)*

We developed a deep learning model to classify the EGMs' classes into normal, abnormal and unclassified. This model integrates Convolutional Neural Network (CNN) and Long Short-term Memory (LSTM) components, having been inspired by previous studies [16], [17] which utilised residual blocks using 1-dimensional convolutional layers, followed by a series of stacked LSTM layers to train the network. For comparison, the CNN-LSTM model's performance was evaluated and compared against a separate machine learning model, the rule-based Decision Trees method (refer to Appendix A for details).

The input to the CNN-LSTM network is a finite-duration discrete signal of $n$ values, expressed as a vector $\bar{x} = [x_1, x_2, \cdots, x_n] \in \Re^n$, where each value denotes the EGM voltage at a particular time instance, and the network outputs a single EGM signal label - normal, abnormal, or unclassified (Fig 3a).

An overview of the architecture is shown in Fig 3a. The network takes as input a time-series of an EGM signal lasting 1500ms with linearly normalised voltage values with a maximum amplitude of 1 (refer to the previous section on data augmentation for details on the processing of the EGM data). The input is passed through a series of modules labelled 'Head', 'Resblock', 'ResBlockSub', and 'Tail' and finally giving the final classified label. 'Skip' (or shortcut) connections are employed within the neural network to help in improving the performance and stability of the model by allowing information to propagate well through the network. The details of the individual modules are given in Appendix B.

We explored 2 variants of the CNN-LSTM network in this study. The first variant is single-branched, taking in only the normalised EGM signal as the input (Fig 3a) as described, while the second variant is a double-branched network that takes in the normalised EGM signal as well as the power spectrum representation of its Fast Fourier Transform across 1500 discrete frequencies (Fig 3b). This value of 1500 frequencies is chosen, as it has the same length as the input EGM samples so that the same model architecture can be reused to process the FFT output, while only requiring a merging operation (concatenation) at the end of the network to combine the features extracted from the EGM and the FFT. Ultimately, the output from the branching in both variants are passed to the 'Tail' module for prediction (Fig. 3b). Notably, within the 'Tail' module, another significant difference arise from the shape of the tensor passed into the LSTM/Global Average component - for the single-branched variant, tensor is of shape $(B, int(1500/2N), 64 * N)$, where B is the batch size, N is the number of stages in our network (as previously discussed). For the double-branched variant, as the outputs from the two branches are concatenated along the last axis before being fed into the tail module, the shape of the input is $(B, int(1500/2N), 64 * N)$.

For model training, at every epoch, the training data was oversampled using random cropping to ensure balanced data in all categories. The network was trained for 100 epochs with the Adam optimizer parameterised as follows: learning rate=$10^{-3}$, β1=0.9, β2=0.999 [18]. The batch size was set to be 32. All computations were carried out on a Linux workstation with Intel Core i7-4790 CPU with 3.6 GHz clock speed, 16 GB RAM and a GeForce GTX TITAN X. It took approximately 9 min for one epoch and a total of 30 hours to train the model using the above-mentioned workstation. Tensorflow 2.4.0 implementation was used in our study. The parameters which yielded the highest validation accuracy among the 100 epochs are saved and used for testing.

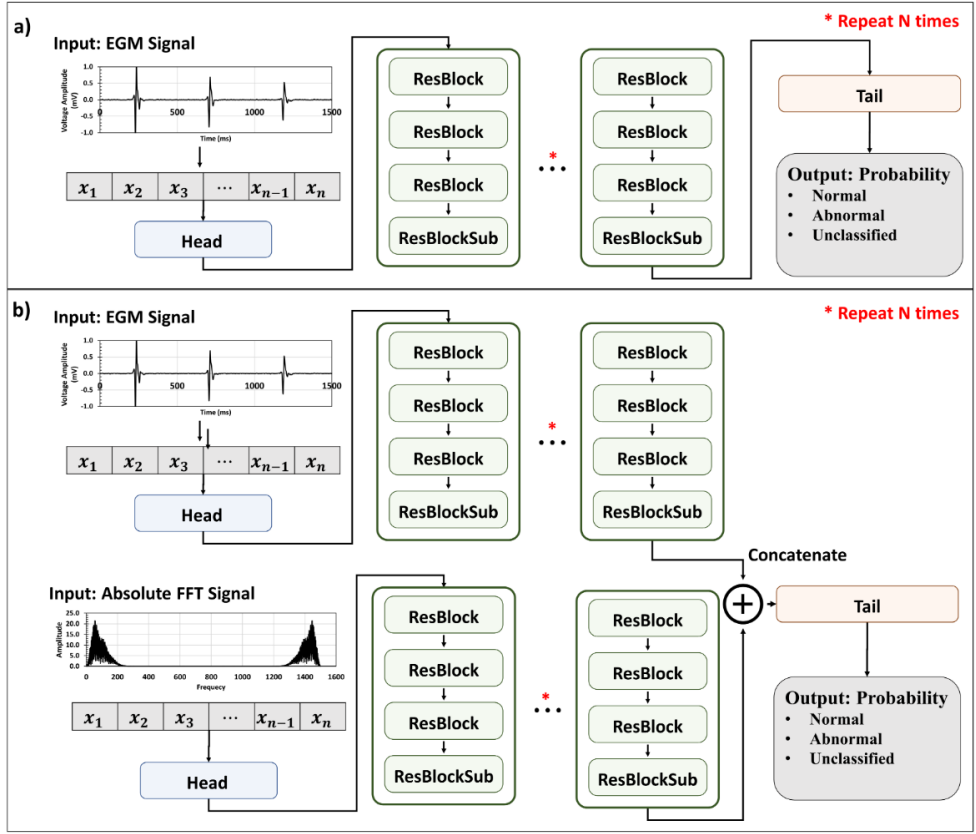
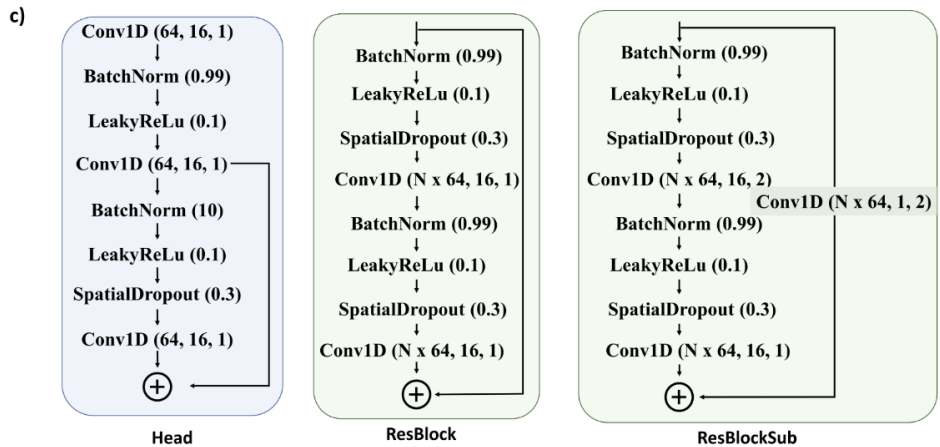
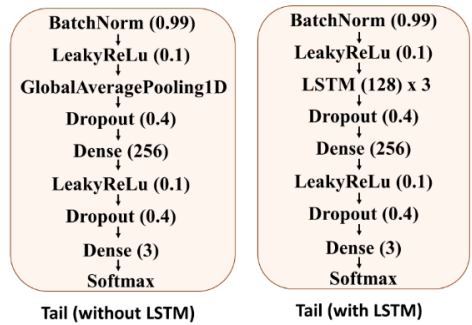

*Figure 3.* Overview of two neural network architectures used (divided into Head, ResBlocks, ResBlockSub and Tail. (a) A normalised EGM is taken as input and a confidence score is assigned to each of the three classes, indicating how likely the EGM is of that class. (b) Both normalised EGM and its Fast Fourier Transform (FFT) were fed into the models and concatenated the features after N-Stage for classification. (c) Details architectures for Head, ResBlock, ResBlockSub, Tail (without LSTM) and Tail (with LSTM).

# Results

## Adopted Data Annotation Using Unanimous Labels

Table 3 shows the confusion matrix of the three researchers' unanimous labels against the cardiac specialists' labels in classifying EGM signals into 'normal', 'abnormal' or 'unclassified'. These labels were performed on a subset of the testing dataset, comprising a total of 135 EGM signals. The researchers' unanimous labels achieved an average of 90% accuracy (one-vs-all), suggesting that the data labelling performed by researchers is of sufficient quality for training.

*Table 2.* Confusion matrix of unanimous labels against the cardiac specialists' labels. Results showed that our unanimous labels are 90% accurate.

|  | Types (Number of Signals) | Researchers' Unanimous Labels on subset of the test dataset | | |
|---|---|---|---|---|
|  |  | **Normal** | **Abnormal** | **Unclassified** |
| **Cardiac Specialists' labels** | Normal (60) | 50 | 2 | 8 |
|  | Abnormal (37) | 2 | 34 | 1 |
|  | Unclassified (38) | 0 | 0 | 38 |
| **Performance (One-vs-all)** | Precision | 0.96 | 0.94 | 0.81 |
|  | Recall | 0.83 | 0.92 | 1.00 |
|  | **F1** | 0.89 | 0.93 | 0.8 |

|  | Accuracy | 0.90 |
|---|---|---|

## Performance of Convolutional Neural Network, Long Short-term Memory (CNN-LSTM) Model

Table 3 shows the validation accuracies of each of the four CNN-LSTM model variants (shown in the columns), which are combinations of networks with and without LSTM components, and using either normalised EGM only, or normalised EGM with its FFT. Comparisons are also made between the different numbers of residual blocks (N) used in each model variant, as shown in the rows of Table 3. All model variants showed comparably high predictive performance, with validation accuracy ranging from 0.88 to 0.94 for all variants. No specific trend was observed in the validation accuracy when the number of residual blocks was increased. The introduction of the additional FFT input actually led to a degradation in performance. This may be due to the FFT output having drastically different scales of magnitudes from the EGM signals (which have amplitudes normalised to 1), causing instability in training.

*Table 3. Validation accuracies attained by two different models - (i) Model with EGM as input and (ii) Model with EGM and FFT as inputs. N refers to the number of residual blocks. The highest validation accuracy for each particular model is shown in bold.*

|  | Inputs - EGM |  | Inputs - EGM and FFT |  |
|---|---|---|---|---|
| N | With LSTM | Without LSTM | With LSTM | Without LSTM |
| 1 | 0.87 | 0.90 | 0.89 | 0.91 |
| 2 | 0.92 | 0.89 | 0.92 | 0.93 |
| 3 | 0.89 | 0.90 | 0.93 | 0.93 |
| 4 | 0.92 | 0.92 | 0.91 | 0.90 |
| 5 | 0.93 | 0.92 | 0.92 | 0.91 |
| 6 | **0.94** | 0.90 | **0.94** | 0.90 |
| 7 | 0.91 | 0.91 | 0.93 | 0.91 |
| 8 | 0.88 | 0.90 | 0.93 | 0.90 |
| *Mean* | *0.92* | *0.91* | *0.92* | *0.91* |
| *Std dev* | *0.02* | *0.01* | *0.02* | *0.01* |

Table 4 shows the classification results of applying the CNN-LSTM on the testing dataset. The results from the Decision Trees Model on the same dataset are also included in Table 4, and further details about the Decision Trees Model's performance can be found in Appendix C. For the CNN-LSTM models, results from the two model variants with the highest validation accuracies (from Table 3) were chosen to be presented - (i) Model 1: 6-Resblock LSTM Model and EGM input and (ii) Model 2: 6-Resblock LSTM Model with EGM and FFT inputs. These models' confusion matrices are shown in Table 4. Model 1 with 6-Resblock LSTM model with EGM as input achieved a testing accuracy of 0.81. Model 1 is able to predict normal, abnormal and unclassified well with high precision, recall and F1-score. All evaluation metrics (precision, recall and F1-score) achieved a score of at least 0.74. Model 2, with 6-Resblock LSTM model using EGM and FFT as inputs, predicted unclassified perfectly with a recall of 1.00 but the predictive performance on normal and abnormal signals were poor with a recall of 0.52 and 0.48, respectively. The F1-score for each signal category was approximately 0.61.

From Table 4, in comparison, for the Decision Trees algorithm, it can be seen that the performance of the decision tree model in classifying 'unclassified'-labelled signals has the highest recall (1.00) but the lowest precision (0.56). The model performed with a precision of 0.63 and a recall of 0.67 for 'abnormal' signals, and has high precision (0.93) but low recall (0.51) for 'normal' signals. The model has the highest F1-score (0.72) for 'unclassified' signals, followed by 'normal' signals (0.66) and lastly 'abnormal' signals (0.65). The overall average testing accuracy is 0.67. It is also observed that 50% of normal signals were mis-classified as 'abnormal' and 'unclassified' while 29% of 'abnormal' signals were mis-classifed as 'unclassified'. All 'unclassified' signals were predicted correctly.

Table 4. Confusion matrix of testing dataset using two models - (i) Model 1: 6-Resblock LSTM model, with EGM as input and (ii) Model 2: 6-Resblock LSTM model, with EGM & FFT as input. The model with higher testing accuracy is Model 1, achieving an accuracy of 0.81.

| | Types (Number of Signals) | Model 1: 6-Resblock LSTM model, with EGM as input | | | Model 2: 6-Resblock LSTM model, with EGM & FFT as inputs | | | Decision Trees Model | | |
| --- | --- | --- | --- | --- | --- | --- | --- | --- | --- | --- |
| | | Prediction | | | Prediction | | | Prediction | | |
| | | Normal | Abnormal | Un-classified | Normal | Abnormal | Un-classified | Normal | Abnormal | Un-classified |
| Label | Normal (83) | 66 | 10 | 7 | 43 | 27 | 13 | 42 | 21 | 20 |
| | Abnormal (52) | 8 | 39 | 5 | 0 | 25 | 27 | 3 | 35 | 14 |
| | Unclassi-fied (43) | 0 | 4 | 39 | 0 | 0 | 43 | 0 | 0 | 43 |
| Precision | | 0.89 | 0.74 | 0.76 | 1.00 | 0.48 | 0.52 | 0.93 | 0.63 | 0.56 |
| Recall | | 0.80 | 0.75 | 0.91 | 0.52 | 0.48 | 1.00 | 0.51 | 0.67 | 1 |
| F1 | | 0.84 | 0.74 | 0.83 | 0.68 | 0.48 | 0.68 | 0.66 | 0.65 | 0.72 |
| Accuracy | | **0.81** | | | 0.62 | | | 0.67 | | |

Figure 4 shows the representative signals that were classified wrongly by Model 1, the best-performing model in terms of classification accuracy. It was observed that abnormal cases that were misclassified as 'normal' exhibited normal EGM's characteristics - clear and distinctive periodic activations. For normal EGMs that were mis-classified as 'abnormal' signals, the model categorises signals with extra activation and slightly long activation into abnormal signals. Lastly, signals that have very low amplitude were classified as 'unclassified' by the model. All these cases seem to be examples of difficult cases which require the neighbouring signals' information to obtain the correct categories.

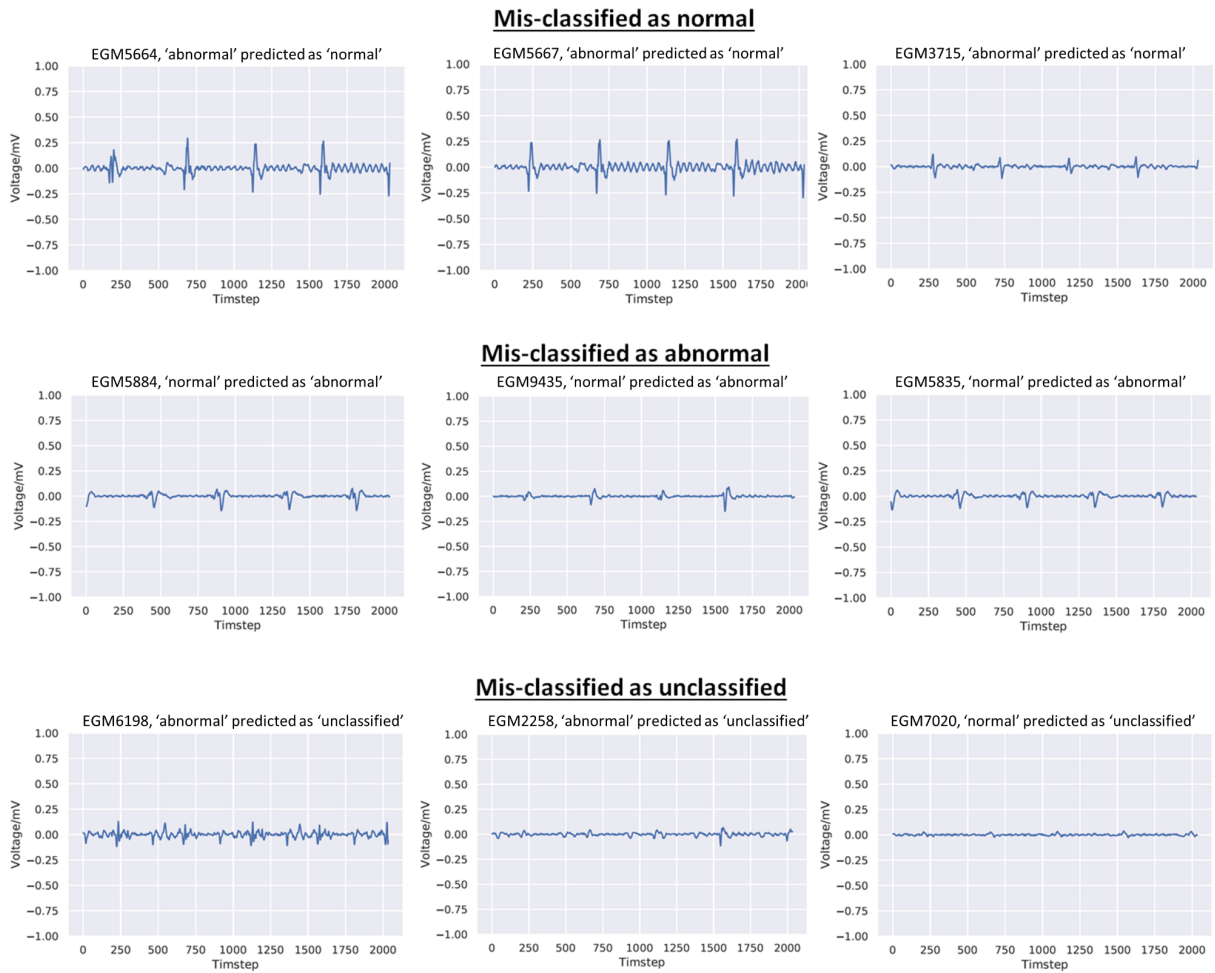

*Figure 4.* Representative EGM signals in testing datasets that were wrongly classified by the deep learning model with Model 1 - 6-Resblock LSTM model, with EGM as input. Each graph's title shows a EGM identification number, and the type of misclassifications.

## Discussion

In this study, we developed the Convolutional Neural Network with Long Short-term Memory component, for the classification of intracardiac EGMs gathered from patients diagnosed with tachycardia. This forms a crucial step towards the identification of the isthmus for ablation in atrial tachycardia treatment.

The identification of the ablation location in reentrant tachycardia requires an accurate activation map for precise map interpretation. One of our main contributions is that our proposed ML models do not require LAT information for prediction, which eliminates the errors and biases which may be introduced by humans producing these LAT annotations. Our model is trained using only the EGMs recordings as inputs to predict different types of EGMs signals. We classified EGM into three categories: (i) normal, (ii) abnormal and (iii) unclassified. Abnormal signals comprise fractionated signals that are commonly seen near reentrant point, slow conduction region and line of block, which could be potential locations for ablation [10].

From our results, we observe that the testing accuracy is significantly higher using the CNN-LSTM instead of the Decision Trees model - the testing accuracies were 0.81 and 0.62, respectively. This finding indicates that analysing EGM signals using a single set of explicitly specified rules is not suitable as EGM signals are complex. Deep learning, on the other hand, has the ability to learn the complex, intrinsic features within the signals and identify useful features to differentiate the EGM signals, achieving a strong testing accuracy of 81%. Similar observations were also reported where deep learning models outperformed classic rule-based classifiers that use classical analyses of cycle length, dominant frequency and autocorrelation handcrafted features to predict sinus rhythm or atrial arrhythmias [19]. This showed that although the decision trees model is advantageous in providing transparency and interpretability in the classification process, it is found to be less robust than deep learning approaches.

Using a small dataset (820 bipolar EGM signals from 9 cases), the CNN-LSTM achieved a comparable accuracy to the model presented by Liao et al. (testing accuracy = 87.9%), which was trained on a total of ~13k unipolar EGM recordings [20]. Zolotarev et al., similarly achieved an F1 score of 0.81 using 28k unipolar EGM recordings to predict ablation location [21]. This could be due to unipolar signals having increased susceptibility to noise [22] that contribute redundant information to the model and thus more recordings are needed.

In our experiments, we found that including the FFT information from the EGM to the CNN-LSTM did not lead to an improvement in model performance. This finding is well-corroborated with a previous study where they found that dominant frequency, computed from the FFT analysis, was the 7th top feature of out 11 features used. The overall accuracy of their Support Vector Machines model did not improve much after adding the dominant frequency feature into their model [23].

Figure 4 plots the representative EGMs signals that were classified wrongly by the CNN-LSTM. It can be seen that the cases that were classified wrongly have characteristics that belong to other classes. For example, the abnormal EGM signals that were classified as 'normal' exhibit normal EGM's characteristics, which are distinctive activation with clear periodicity, while signals that were classified as 'unclassified' have either noisy periodic pattern or subtle low voltage activation. These ambiguous signals would be a problem for the current LAT annotation mapping procedure, where the activation timing is not clear. Incorporating neighbouring EGM signals into the model may mitigate this problem.

There are limitations in our study. Firstly, because of the relatively small size of our dataset, it may not be diverse enough to train a well-generalised model, especially because some characteristics of abnormal EGMs may differ between patients. Secondly, in this study, we combined different kinds of abnormal EGMs into a single 'abnormal' class, and this meant that information pertaining to the specific subcategories of abnormal signals, such as line of block or slow conduction, are not emphasised. This was a tradeoff undertaken as the number of samples would have been reduced significantly if we had chosen to create more categories to address each abnormal subtype. It is known that deep learning is a data-hungry method and data annotation in this complex tachycardia problem is expensive. It remains as a future work to obtain more clinical data so that the abnormal category can be further stratified.

## Conclusions

We have developed two novel machine learning methods for the classification of intracardiac EGM signals encountered in atrial tachycardia - the Decision Trees model, and the CNN-LSTM. Our results showed that the 6-Resblock with LSTM model using EGM recordings as inputs, is the best model that can achieve a high accuracy of 81%, when benchmarked against other works in literature. We also demonstrated that the CNN-LSTM has a superior performance than Decision Trees.

# Acknowledgments

The authors wish to thank Mr Isaac Cheong and Mr Davide Coppola for their helpful advice and comments on the paper.

# Funding

Biomedical Research Council (BMRC); National Heart Centre of Singapore (Nurturing Clinician Scientist Scheme (NCSS) Grant 07/FY2019/P1/16-A31)

# Appendix

## *A. Decision Trees Model*

The Decision Trees model utilises a set of decision rules for the classification of EGM signals. Each EGM signal can be thought of as a time series of voltage values acquired at a specific spatial location, and comprises sequential activation cycles in which the instances of activation are marked by signal impulses (Fig A1a). Signal preprocessing of the EGM is first performed, followed by its classification using decision trees. Broadly, this can be broken down into three steps:

1. take the absolute value of the entire EGM signal, which makes the signal positive-only, and identify the point with the highest voltage value within the signal - this point is our highest activation peak

2. Find all the other activation peaks in the EGM signal based on a margin (± 5%) of a given cycle length. The cycle length is given by the patient's ECG taken in conjunction with the acquisition of intracardiac EGMs, where the average value of the ECG cycle length of all 9 patients is 606 ± 227 ms. The highest activation peak is used as a reference to identify the other peaks. The time instance of the highest peak was also obtained in the first step. By defining a window width of a proportion (eg. 10%) of the cycle length centred at this time instance, the window is strided in both directions along the EGM signal's time axis by the stride length (eg. 95% of the cycle length) to identify the other activation peaks (Fig A1c). These example values give a margin of ± 5% of the entire cycle length. The window width is a hyperparameter of the model as well as the amount of padding applied to the window width, known as the window padding.

3. Divide the EGM signal into regions bounded in time by the time instances of the activation peaks. Each region is bounded by two consecutive activation peaks. Perform analysis on every region. Here, a two-step process, using a decision tree at each step, is used to classify the entire EGM signal.

The first set of decision trees within the overall model determines the label of the region (Fig A1e). Two thresholds are defined based on the voltage amplitude of the highest peak within the EGM: the 'unclassified' and 'abnormal' threshold values. There are two activation peaks bounded within each region:

Classification of each regions in the whole EGM

- If any of the two activation peaks has a voltage value that is higher than the unclassified threshold value (defined as 15% of the highest peak), the region will be labelled as 'unclassified'.
- Otherwise, it was further evaluated if both peaks are below the abnormal threshold value (defined as 10% of the highest peak). If this is so, the region will be labelled as 'normal'.
- Or else, if either peak or both peaks are above the abnormal threshold (defined as 10% of the highest peak), the region will be labelled 'abnormal'.

Classification of whole EGM signal by aggregating the classification of each region

The second set of decision rules was used to determine the classification of the entire EGM signal by aggregating the classification results from all the regions using the rules shown in Figure A1f. If the EGM:

- has less than 3 activation cycle peaks or if at least one of its component regions are labelled as 'unclassified', the EGM sample will be labelled as 'unclassified';
- if all regions are labelled as 'normal', the EGM signal will be classified as 'normal';
- if all regions are labelled as 'abnormal', the EGM will be classified as 'abnormal';
- Lastly, if the EGM sample contains alternating normal and abnormal regions, this indicates a normal EGM obtained near the ventricle, hence the EGM will be classified as 'normal'.

The hyperparameters to the decision tree algorithm are: the search window size and amount of window padding, and thresholds for deciding on 'unclassified' and 'abnormal' signals. To choose the best values for the hyperparameters, we tested different combinations of the hyperparameters using suitable ranges for each parameter using both the training and validation dataset. The resulting hyperparameters which gave the highest (for both training and validation) accuracy were selected. This combination of hyperparameters was then used on the testing dataset to evaluate the performance of the Decision Trees model. We then compared the performance of decision trees with and without augmentation.

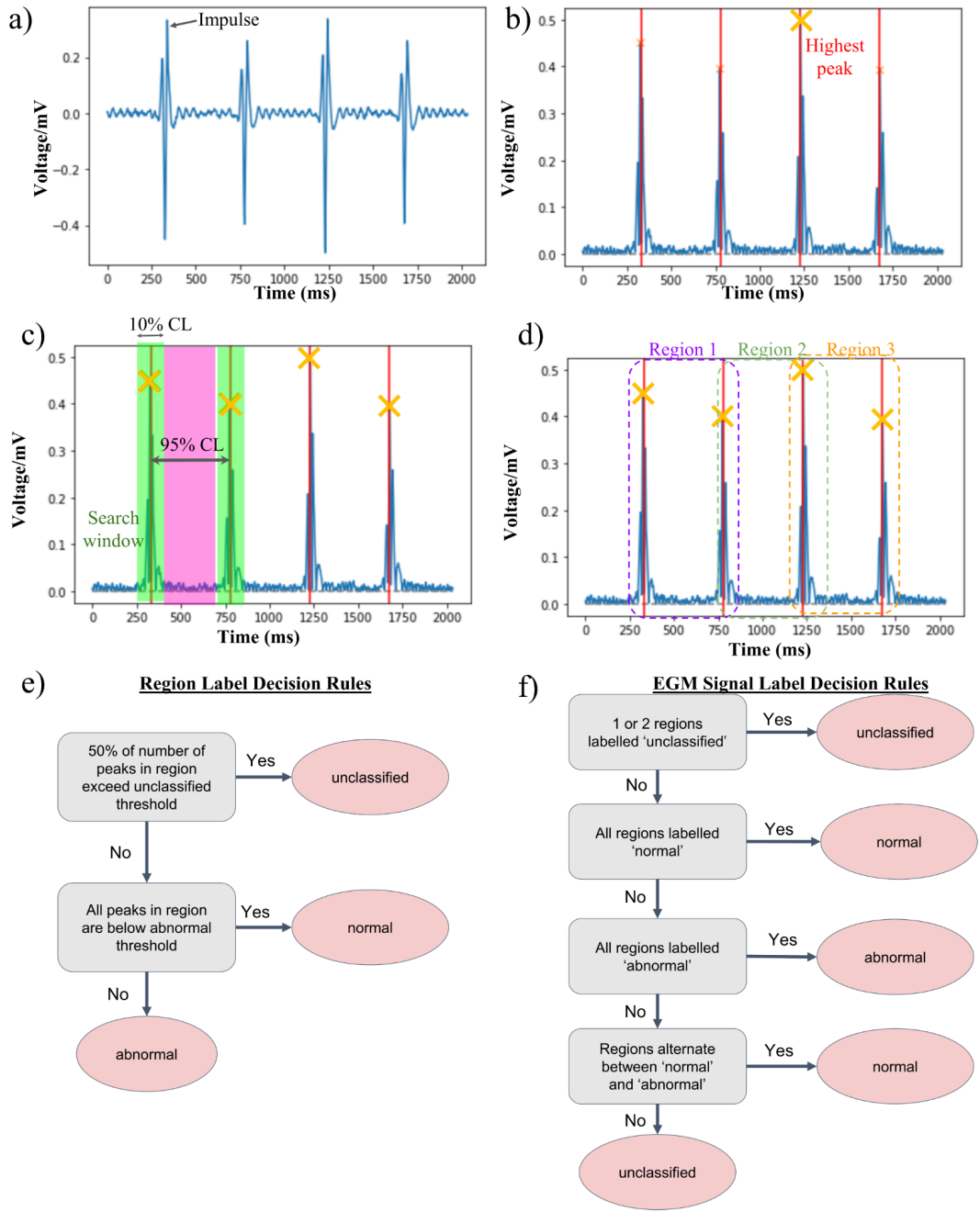

*Figure A1.* Overview of our Decision Trees model. (a) Original EGM. (b) Absolute-valued EGM, where the maximal peak is identified. (c) Search window width and stride length for signal peaks identification. Example values are given in this example for illustration. (d) Defined regions of the EGM signal, where the label of one region of the EGM will be classified using a set of rules shown as a flowchart in Fig. A1e. The same labelling process is applied for all regions in an EGM (3 regions in this sample). (e) Decision rules for determining labels of a single region. (f) Second decision ruleset for determining the final label of the EGM. For this

*sample, the final prediction is normal since all regions are predicted as normal.*

## B. Deep Learning Model: Convolutional Neural Network, Long Short-term Memory (CNN-LSTM) Modules

The 'Head' module (Fig B1) consists of a sequence of 8 layers: a 1D Convolution layer (using 64 filters, kernel size of 16 and a stride of 1), a Batch Normalisation layer, a LeakyReLU activation, a second 1D Convolution layer (using 64 filters, kernel size of 16 and a stride of 1), another Batch Normalisation layer, a LeakyReLU activation, followed by the application of 1D Spatial Dropout (which implement drop-out of entire 1D feature maps) and a final 1D Convolution layer (using 64 filters, kernel size of 16 and a stride of 1). A shortcut connection is introduced in adding the output of the second 1D Convolution layer to the output of the final 1D Convolution layer. This aids model performance by allowing information to propagate through to subsequent modules.

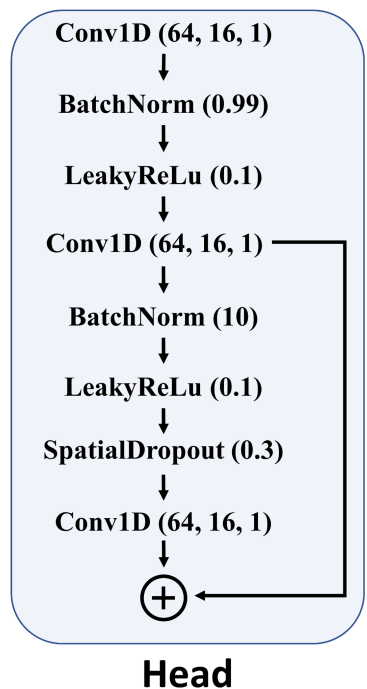

**Head**

***Figure B1.** The 'Head' module*

The output of the 'Head' module is piped to repeated stacks of three sequential 'ResBlock' modules (Fig B2) followed by a 'ResBlockSub' module (Fig B3), as seen in Figure 3a (Methods section). A variable number of stacks (N) was tested out in our study, ranging from 1 to 8. Subsequently, the output of the repeated stacks of 'ResBlock' and 'ResBlockSub' are output to the 'Tail' module. Thus far, the workhorse of the network has been the 1D Convolution layers, and the LSTM component of the network is ultimately introduced here in the 'Tail' module. The introduction of the LSTM here confers the network the ability to learn complex temporal patterns spanning the entire signal based on the local features produced by the preceding convolution layers.

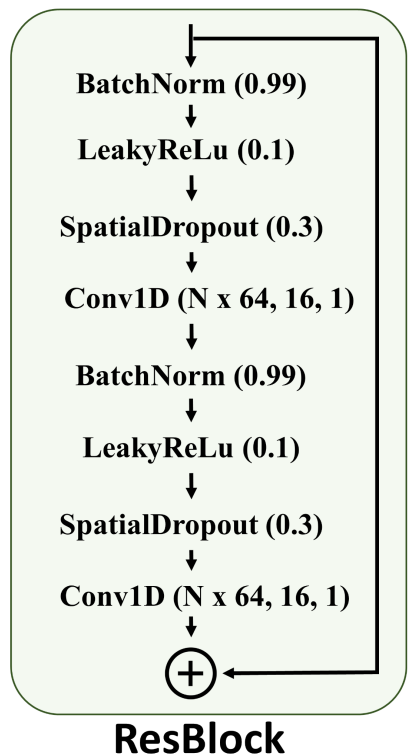

**Figure B2.** *The 'ResBlock' Module*

Within each 'ResBlock' module (Fig B2) are 8 layers: a Batch Normalisation layer, a LeakyReLU activation, followed by application 1D Spatial Dropout, a 1D

Convolution layer (using N times 64 filters, kernel size of 16 and a stride of 1), followed by a second Batch Normalisation layer, a second LeakyReLU activation, second application of 1D Spatial Dropout, and a final 1D Convolution layer (using N times 64 filters, kernel size of 16 and a stride of 1). A shortcut connection is introduced by adding the output of the preceding module to the output of the final convolution layer.

The 'ResBlockSub' module (Fig B3) is identical to the 'ResBlock' module, except that it downsamples its input by a factor of 2 through strided convolutions. This is achieved through the first 1D Convolution layer, which has N times 64 filters, using a kernel size of 16 and a stride of 2. Correspondingly, in the shortcut path a 1D Convolution layer (N times 64 filters, kernel size of 1 and a stride of 2) is introduced to sample its inputs to match the shape of the output from the main branch, in order for them to be added.

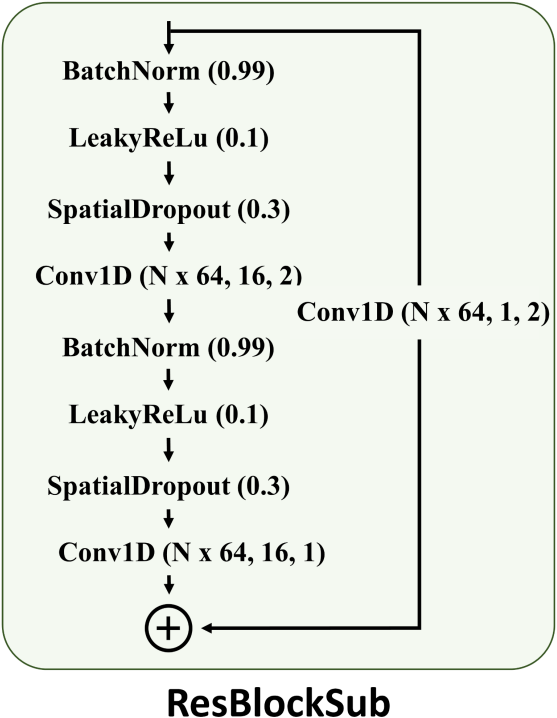

**ResBlockSub**

***Figure B3.*** *The 'ResBlockSub' Module*

The 'Tail' module has 9 layers joined end-to-end. There are 2 variations of 'Tail' modules explored in our study (Fig B4), with the only difference being the choice of using a 1D Global Average Pooling layer instead of the LSTM component. The modules present are: a Batch Normalisation layer, followed by a Leaky ReLU activation, either an LSTM component, comprising of 3 stacked LSTM layers joined end-to-end, with each LSTM layer each consisting of 128 cells - the dimensionality of the input signal, or a Global Average Pooling layer, application of Dropout to random individual inputs, into a fully connected layer of 256 hidden nodes into a LeakyReLU activation, to application of Dropout followed by a Dense layer into a logits vector of length 3, and finally a Softmax layer to output the predicted probability distribution over the 3 labels for the EGM signal. The input to the LSTM layers (after the leaky ReLU activation) consists of feature vectors, which are passed in sequentially, through each LSTM stack as shown in Fig B5.

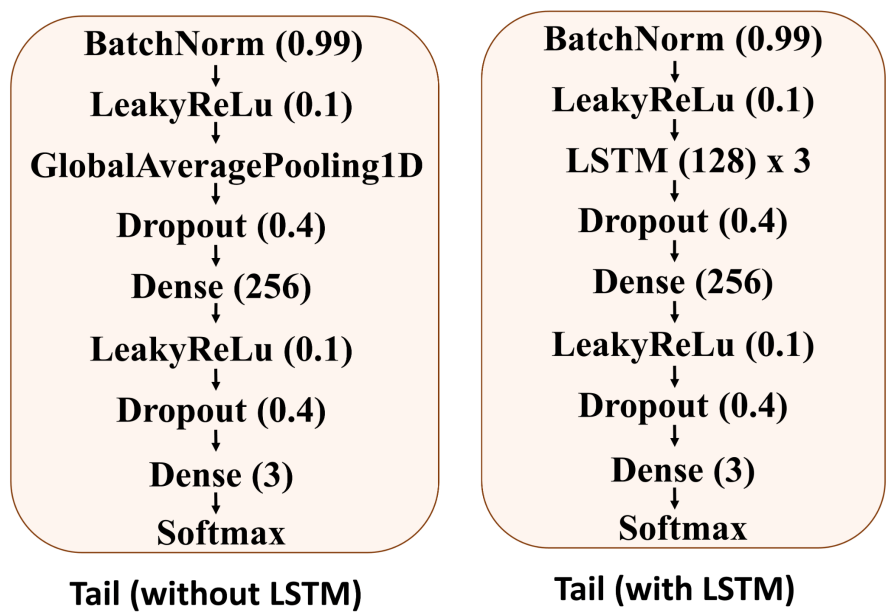

*Figure B4.* Two variants of the 'Tail' module

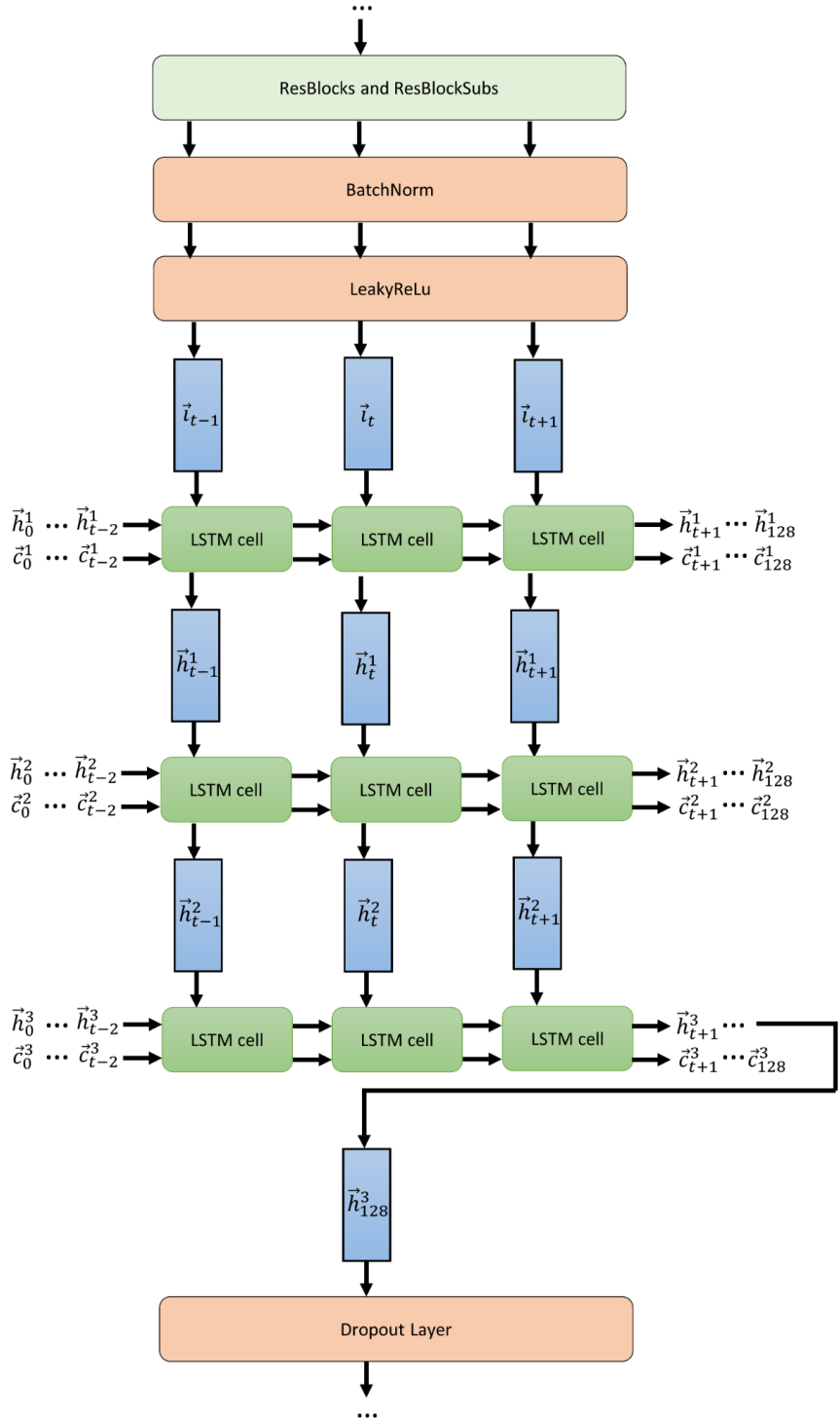

*Figure B5.* Illustration of how the output vector (of length 128, for example) from the leaky ReLU activation layer (preceding the LSTM layer) is passed into the stacked LSTM layers. The input feature vector is denoted by $i_t$, with each hidden state and cell state denoted by $h_t^j$ and $c_t^j$ where $j$ denotes the LSTM layer. The first hidden state and cell states (i.e.

*at t = 0, $h_0^j$ and $c_0^j$) are zero vectors.*

## C. Performance of the Decision Trees Model

Table C1 shows the Decision Trees model's training and testing accuracies for two types of data augmentation on the EGM signals - normalisation and random cropping, as described in the Methods section. The highest testing accuracy achieved was 72.8%, using the normalised EGM signal and without using the cropping augmentation.

The effect of cropping the EGM signals was observed to decrease accuracies across the board. The training accuracy dropped from 71.6% to 69.4% and from 72.8% to 70.2% when cropping was introduced to non-normalized and normalised EGMs, respectively. Likewise, testing accuracy also dropped from 56.7% to 56.2% for non-normalized EGM and 67.4% to 64.6% for normalised EGM when cropping was introduced. This is possibly due to the cropping out of EGM cycle peaks in several EGM signals during the random cropping process, resulting in the total number of cycle peaks being less than three and being labelled as unclassified signals erroneously. As such, if the actual labels of signals being labelled are 'normal' or 'abnormal', the algorithm would have mislabelled them.

On the other hand, the effect of normalising the EGM resulted in higher training and testing accuracies. There is an increase of accuracy from 71.6% to 72.8%, and from 56.7% to 67.4% for training and testing sets (without cropping for both), respectively.

The combination of normalisation and no cropping of the EGM constitutes the augmentations resulting in the best training and testing accuracies. The confusion matrix of the use of the Decision Trees model on the test dataset with normalised and uncropped EGM as input is shown in Table 5 (Results section). The hyperparameters used in training the decision tree model were: a window size of 0.40 (proportional to cycle length of a given patient), an 'unclassified' label threshold of 0.15 (proportional to the highest peak value in a given signal), an 'abnormal' threshold of 0.10 (proportional to the highest peak in a given signal), and a window padding of 45 (in the same units as the timestep of a given EGM monitoring device). From Table 5, it can be seen that the performance of the decision tree model in classifying 'unclassified'-labelled signals has the highest recall (1.00) but the lowest precision (0.56). The model performed with a precision of 0.63 and a recall of 0.67 for 'abnormal' signals, and has high precision (0.93) but low recall (0.51) for 'normal' signals. The model has the highest F1-score (0.72) for 'unclassified' signals, followed by 'normal' signals (0.66) and lastly 'abnormal' signals (0.65). The overall average testing accuracy is 0.67. It is also observed that 50% of normal signals were mis-classified as 'abnormal' and 'unclassified' while 29% of 'abnormal' signals were mis-classifed as 'unclassified'. All 'unclassified' signals were predicted correctly.

*Table C1. Training/Test accuracies of the decision tree to determine the effectiveness of input normalisation (Normalisation) and augmentation by random cropping (Cropping).*

| Accuracy | With Normalisation | | Without Normalisation | |
|---|---|---|---|---|
| | With Cropping | Without Cropping | With Cropping | Without Cropping |
| Training + Validation | 70.2% | 72.8% | 69.4% | 71.6% |
| Test | 64.6% | 67.4% | 56.2% | 56.7% |